\documentclass[aps,prd,onecolumn,showpacs,nofootinbib,amsmath,amssymb,amsfonts,showkeys]
{revtex4}
\usepackage{graphicx}
\usepackage{bm}
\usepackage{xcolor}

\begin{document}
\title{White dwarfs as a probe of dark energy}

\author{S. Smerechynskyi$^1$}
\email{sviatoslav.smerechynskyi@lnu.edu.ua}
\author{M. Tsizh$^1$}
\author{B. Novosyadlyj$^{1,2}$}
\affiliation{$^1$Ivan Franko National University of Lviv, \\ Kyryla and Methodia St., 8, Lviv, 79005, Ukraine}
\affiliation{$^2$College of Physics and International Center of Future Science of Jilin University, Qianjin St. 2699, Changchun, 130012, R.China}
\date{\today}

\begin{abstract}
We investigate the radial density distribution of the dynamical dark energy inside the white dwarfs (WDs) and its possible impact on their intrinsic structure. The minimally-coupled dark energy with barotropic equation of state which has three free parameters (density, equation of state and effective sound speed) is used. We analyse how such dark energy affects the mass-radius relation for the WDs because of its contribution to the joint gravitational potential of the system. For this we use Chandrasekhar model of the WDs, where model parameters are the parameter of the chemical composition and the relativistic parameter. To evaluate the dark energy distribution inside a WD we solve the conservation equation in the spherical static metric. Obtained distribution is used to find the parameters of dark energy for which the deviation from the Chandrasekhar model mass-radius relation become non-negligible. We conclude also, that the absence of observational evidence for existence of WDs with untypical intrinsic structure (mass-radius relation) gives us lower limit for the value of effective sound speed of dark energy $c_s^2 \gtrsim 10^{-4}$ (in units of speed of light). 
\end{abstract}
\maketitle
\section{Introduction}
The nature of dark energy, a substance which causes the observable accelerated expansion of Universe, has become highly studied subject in cosmology in the last two decades. A significant part of models that explain it are (usually, scalar) field models of the dark energy \cite{Ratra88, Turner07, Ellis08, Caldwell09, Amendola2010,Blanchard2010,Wolschin2010,Novosyadlyj2015}. Unlike cosmological constant, scalar field dark energy is assumed to be dynamical and perturbable, changing its density across the time and space, and having as a result restrained impact on the evolution of the large scale structures \cite{Tsizh2015}. Such dark energy can be modelled as the perfect or imperfect fluid, which is effectively described by hydrodynamical parameters: the density $\rho_{de}$, the equation of state parameter $w_{de}$ and the effective speed of sound $c_s$. The most conservative models assume that only density of the dark energy varies with cosmological time, while models with more degrees of freedom assume that all three parameters are dynamical. Depending on its properties dark energy is referred to as the quintessence ($w_{de}>-1$), the phantom ($w_{de}<-1$) or the quintom, in which $w_{de}$ changes its sing during the evolution. Current observable data doesn't give strong preference to any of these types of dark energy \cite{Novosyadlyj2015, Tsizh2015, Sergijenko2014, Planck18}.

It was first noticed by Babichev and co-authors \cite{Babichev2004}, that scalar field dark energy can influence the compact objects through accretion. They analysed how infall of the phantom dark energy "screens" a black hole's (BH) gravitational field, eventually leading even to its disappearing.  

The idea, that hidden components of the Universe can influence the compact objects, like BHs or white dwarfs (WDs), through gravity has developed further. In the last decade, a number of works appeared, in which alternative theories of gravity, also are capable to explain the accelerated expansion of the Universe, were tested on deviation from the general relativity at small scales through impact on the observable features of compact objects. For example, Vainshtein mechanism \cite{vainshtein_1972, babichev_2013}, which restores equivalence of Einstein's general relativity and some of alternative gravity theories at small (Solar System) scales, can be broken inside matter for some cases, such as beyond Horndeski models \cite{Horndeski74, kobayashi_2015, babichev_2016}. With the purpose of probing this scenario different kind of compact astrophysical objects were chosen. For instance, in work \cite{sakstein_2015} the red and brown dwarfs were used as probes for the modified gravity theories through impact on the mass-radius relation, the Chandrasekhar mass limit and the mass-radius relation for the WDs were used in works \cite{jain_2016, banerjee_2017} in order to obtain independent constraints on the Vainshtein breaking parameter. Similar work \cite{babichev_2016} was devoted to the study of relativistic objects such as WDs and neutron stars in which it was shown the importance of post-Newtonian corrections in equilibrium equation for WDs while calculating macroscopic characteristics in the frame of the theory of modified gravity. In the work \cite{das_2015} its authors attempted to explain existence of sub- and super-Chandrasekhar WDs as possible progenitors of peculiar Super Novae of Ia type with help of modified gravity theory. BHs and relativistic stars in scalar-tensor theories of gravity are studied in \cite{Chagoya18, Lehebel18}. In both works authors find constraints from observing compact objects on theories. WDs are used for similar purpose in papers \cite{Saltas18} and \cite{Nari18}: the corrections to equations that describe them are evaluated and constraints on the possible modifications of gravity are given.

The goal of this paper is to investigate the radial density distribution of the dynamical dark energy inside the WDs and estimate its possible impact on their intrinsic structure. It will give the possibility to estimate the lower limit for the value of the effective sound speed $c_s$ of the dark component. We do it by considering static solution of dark energy distribution in spherically symmetrical metric. We take it into account in the equation of Chandrasekhar model, the dark energy changes joint gravitational potential of the system "white dwarf + dark energy", hence changing mass--radius relation for WDs, depending on its parameters, in particular on the effective speed of sound. 

Though local behaviour of dark energy clustering is an object of study in a lot of works lately (see \cite{Donnari2016, Dhawan2017} and \cite{novosyadlyj_2016} for example), and there are even examples of "mixed star" (dark energy + baryon matter) solutions \cite{Chan2008, Stoytcho2011}, our work, as of our knowledge, is the first attempt to constrain dark energy parameters through observable WDs properties. 

The paper is organized as follows. In Section~\ref{sect_2} we briefly remind the Chandrasekhar model of WD and its main results. Section~\ref{sect_3} contains the equation of state for the dark energy and the calculation of its radial distribution inside WD. In Section~\ref{sect_4} we discuss the possibility of setting the constraint on the effective speed of sound of the dark energy using the WDs and in Section~\ref{sect_5} we present our conclusions.

\section{Chandrasekhar model of white dwarfs}
\label{sect_2}

Typical WD is a spherical object with mass of one half that of the Sun and radius of the order of Earth's one. In such extreme dense objects hydrostatic equilibrium is maintained by pressure of relativistic degenerated electron gas \cite{chandrasekhar_1931, chandrasekhar_1935, chandrasekhar_1939}. High density of matter causes the equation of state of electron gas to be almost independent on temperature, that is why mechanical and thermal structure of WDs can be treated separately. Such zero-temperature approximation is especially applicable to massive WDs where finite temperature effects are negligible \cite{vavrukh_2013}. In approximation of complete degeneration the equation of state of non-interacting relativistic electron gas can be written in the parametric form as follows
\begin{eqnarray}
\label{eq1}
&& P_e (r) = \frac{\pi m_e^4c^5}{3h^3} f(x(r)),\nonumber\\
&&f(x) = 8\int\limits_0^x \frac{y^4dy }{\sqrt{1+y^2}} = x(2x^2-3)\sqrt{1+x^2}+3\ln{(x+\sqrt{1+x^2})};\\
&&\rho(r) = \mu_e m_u\frac{8\pi (m_ec)^3}{3h^3} x^3(r)\nonumber.
\end{eqnarray}
Here $m_e$ is the electron rest mass, $m_u$ stands for atomic mass unit and dimensionless chemical composition parameter $\mu_e$ determines the number of nucleons per free electron for an averaged nucleus in a star (we assumed here $\mu_e = 2$). The dimensionless Fermi momentum of electrons $x = p_F/m_ec$, called relativistic parameter, plays the role of parameter in above mentioned equation of state. 

\begin{figure*}
\begin{minipage}[t]{0.49\textwidth}
\centering\includegraphics[width=1\textwidth]{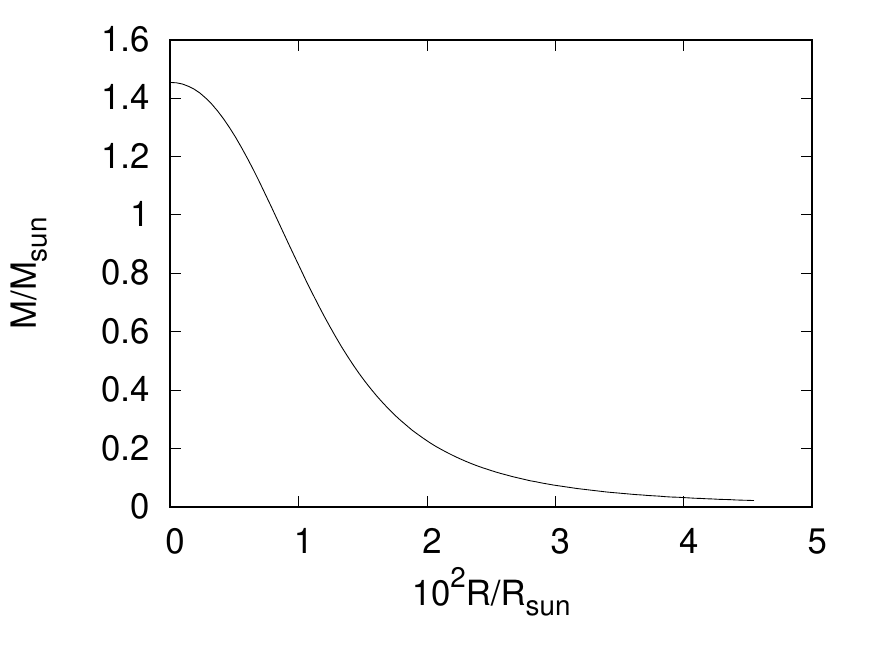}
\end{minipage}
\begin{minipage}[t]{0.49\textwidth}
\centering\includegraphics[width=1\textwidth]{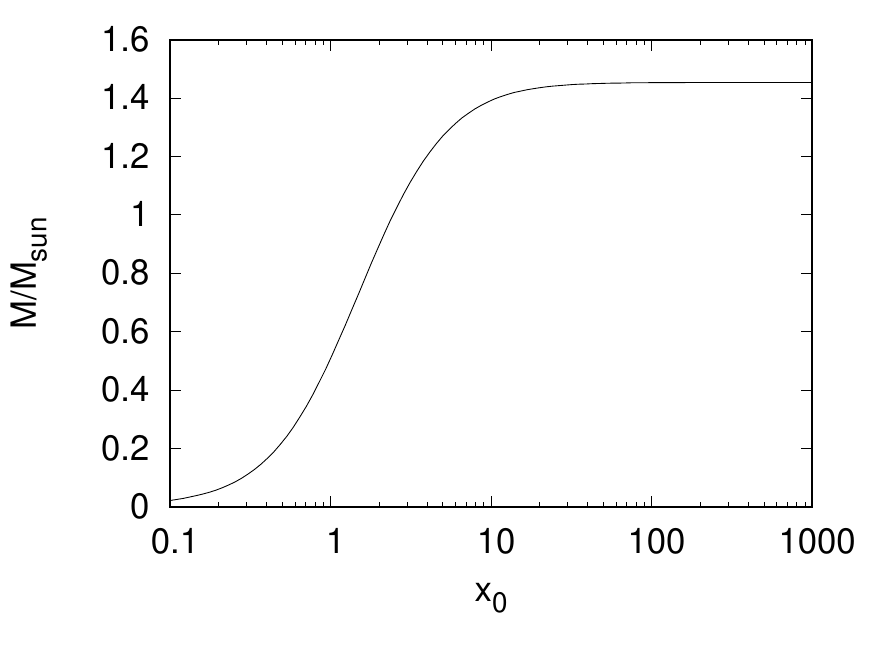}
\end{minipage}
\caption{Characteristics of white dwarfs in Chandrasekhar model:
 the mass-radius relation (left panel), the Chandrasekhar limit (right).}
\label{fig1}
\end{figure*}

Assuming the hydrostatic equilibrium of a non-rotating gaseous sphere Chandrasekhar obtained the model of the WDs with two parameters \cite{chandrasekhar_1939}: relativistic parameter in stellar centre $x_0$ and chemical composition parameter $\mu_e$ (which is close to 2 for all elements except hydrogen). Within this model two important outcomes became famous: peculiar mass-radius relation -- radius of WD decreases with increasing mass, on the contrary to normal stars (left panel of Fig.~\ref{fig1}); and existence of maximum mass of WD $\sim 1.5 M_{\odot}$, known as the Chandrasekhar mass limit. This is formal limit for WD with the central density approaching infinity (right panel of Fig.~\ref{fig1}). The latter one played the crucial role in the discovery of the accelerating expansion of the Universe through observations of the distant supernovae of Ia type. This kind of superluminous events are believed to be explosions of WDs exceeding Chandrasekhar limit due to the accretion of matter from another component of the binary system.

\section{DARK ENERGY in WHITE DWARFS}
\label{sect_3}

\subsection{Dark energy model}
In this work we analyse the scalar field model of dark energy with barotropic equation of state 
\begin{equation}
\label{eq1a}
p_{de}=\omega(\rho_{de})\rho_{de}c^2,
\end{equation}
where $p_{de}$ and $\rho_{de}$ are pressure and density of dark energy, respectively.
We consider model for which relation between the equation of state parameter $w$ and the effective speed of sound $c_s^2$ (in the units of speed of light $c$) is as follows 
\begin{equation}
w=c_s^2-(c_s^2-w_{\infty})\frac{\rho_{\infty}}{\rho_{de}}.\label{w-rho}
\end{equation}
Here $\rho_{\infty}$ is the background density of dark energy (at $r\rightarrow \infty$), for which we adopted the value $10^{-26}$kg/m$^3$ \cite{novosyadlyj_2014}. Also, we considered two types of dark energy, quintessence with $\omega_{\infty}=-0.8$ and phantom with $\omega_{\infty}=-1.2$.

We chose hydrodynamical (phenomenological) representation of scalar field dark energy as usually, for convenience. Scalar field dark energy can be represented as perfect or imperfect fluid with barotropic equation of state (\cite{Sergijenko14}, \cite{Babichev2013}). Indeed, given Lagrangian of the field $\mathcal{L}(X,U)$ with kinetic term $X$ and potential $U$, the connection with phenomenological quantities is as follows
$$\rho_{de} = 2X\mathcal{L}_{,X} - \mathcal{L},\quad p_{de} =  \mathcal{L}, \quad
w_{de} = \frac{p_{de}}{c^2\rho_{de}} = \frac{\mathcal{L}}{2X\mathcal{L}_{,X}}, \quad c_s^2 = \frac{\delta p_{de}}{c^2\delta\rho_{de}} = \frac{\mathcal{L}_{,X}}{2X\mathcal{L}_{,XX} - \mathcal{L}_{,X}} $$
One can obtain considered here linear equation of state for stationary Minkowski or Schwarzschild world and the scalar field dark energy with conditions $c_s^2=const>0$ and $w_{de}<0$ \cite{novosyadlyj_2014}. The properties of such dark energy in the vicinities of compact objects were also studied in \cite{Babichev2004,novosyadlyj_2014,Babichev2013}. In the case of static space-time the EoS parameter (3) corresponds to a static scalar field with a constant potential $U$ and a density-dependent kinetic term $X$ \cite{Novosyadlyj19}.

\subsection{Dark energy distribution inside white dwarf}

In order to analyse the behaviour of dark energy inside a compact astrophysical object we consider the simplest model of WD without rotation and neglect effects of magnetic field, finite temperature and Coulomb interactions on the mechanical structure. Consequently, we expect spherically symmetric distribution of dark energy inside a star. Also, in this work we do not aim to describe the dynamical evolution of dark energy in the gravitational field of compact object, but instead focus on static configuration of system, which consists of two components: matter of WD and dark energy.

The space-time metric for spherically symmetric case can be written in the form
\begin{equation}
\label{eq2}
ds^2=e^{\nu(r)}c^2d\tau^2-e^{\lambda(r)}dr^2-r^2\left(d\theta^2+\sin^2{\theta}d\varphi^2\right).
\end{equation}

In our case the components of metric do not depend on time and can be obtained from Einstein equations with boundary condition $\lambda(r=0) = 0$ 
\begin{eqnarray}
\label{eq3}
&&e^{-\lambda(r)}=1-\frac{8\pi G}{c^2r}\int\limits_0^r \left[\rho_m(r')+\rho_{de}(r')\right]r'^2dr',\\
&&\nu(r)+\lambda(r)=-\frac{8\pi G}{c^2}\int\limits_r^R \left[\rho_m(r')+\rho_{de}(r')+\frac{p_m(r')+p_{de}(r')}{c^2}\right] e^{\lambda(r')}r'dr'.\nonumber
\end{eqnarray}
Here $\rho_m$, $p_m$ are local density and pressure of stellar matter and $\rho_{de}$, $p_{de}$ denote corresponding characteristics of dark energy.

In the case when the equilibrium of the gravitational force and pressure gradient is fulfilled, the following equations hold for both components of considered system
\begin{eqnarray}
\label{eq4}
\frac{dp_m}{dr}+\frac12(\rho_m c^2+p_m)\frac{d\nu}{dr}=0,\\
\frac{dp_{de}}{dr}+\frac12(\rho_{de} c^2+p_{de})\frac{d\nu}{dr}=0.\nonumber
\end{eqnarray}
If the density of dark energy is essentially lower than the matter density and metric function $\nu$ is defined by distribution of matter mainly then the last equation gives the radial distribution of dark energy inside a star  
\cite{novosyadlyj_2014}
\begin{equation}
\label{eq5}
\rho_{de}(r)=\rho_{\infty}\left(\frac{c_s^2-\omega_{\infty}}{1+c_s^2}
    +\frac{1+\omega_{\infty}}{1+c_s^2}\left[e^{\nu(r)}\right]^{-\frac{1+c_s^2}{2c_s^2}}\right).
\end{equation}

In order to solve the system of equations~(\ref{eq3})-(\ref{eq4}) in the general case we have to know the value for $\rho_{de}(0)$ or the value of $\nu(0)$ in stellar centre. For this we applied the iterative procedure: in zero approximation we assumed no influence of dark energy on WD and having the results of Chandrasekhar model (given in (eq.~\ref{eq1})) we calculated second equation in~(\ref{eq3}) at the point $r=0$. Following this, we use found potential $\nu(0)$ to evaluate the density of the dark energy at the center with formula (\ref{eq5}). At the next step this value was used to solve the system~(\ref{eq3})-(\ref{eq4}), where both baryon matter and dark energy are taken into account when calculating the potential, and recalculate a new value of $\rho_{de}(0)$. Such procedure was repeated until the convergence was reached or iteration limit exceeded. The algorithm stops when relative change of the potential at consecutive iterations is less than $10^{-5}$. Usually it takes less then 10 iterations to reach the convergence.

The system ceases to converge when the effective sound speed $c_s^2$ approaches to zero, meaning that dark energy with very small $c_s^2$ doesn't allow static solutions. The specific value of $c_s^2$ when convergence is lost depends on relativistic parameter in the center of the star $x_0$ and this value of $c_s^2$ increases with growth of $x_0$. Technically divergence is manifested through very rapid growth of density and thus mass of dark energy in the system up to the infinity (or negative infinity in the case of phantom dark energy).

\begin{figure}

\begin{minipage}{.32\textwidth}
\includegraphics[width=1\textwidth]{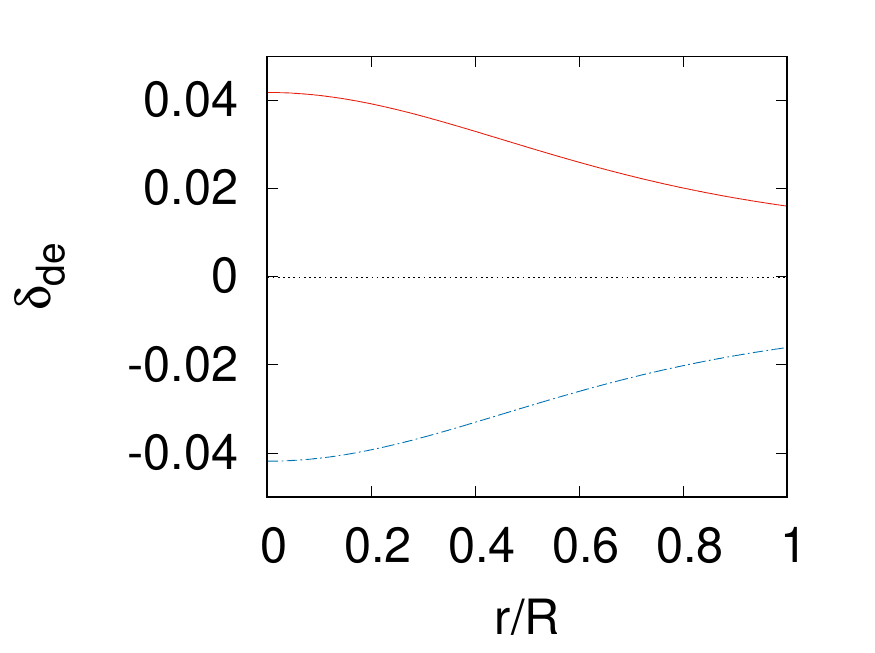}
\centering{a)}
\end{minipage}
\begin{minipage}{.32\textwidth}
\includegraphics[width=1\textwidth]{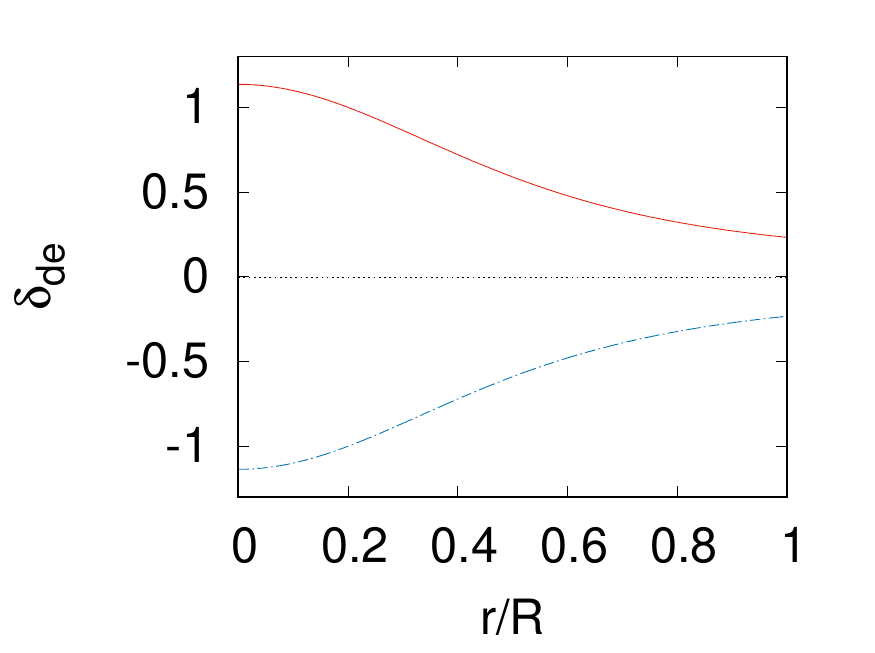}
\centering{b)}
\end{minipage}
\begin{minipage}{.32\textwidth}
\includegraphics[width=1\textwidth]{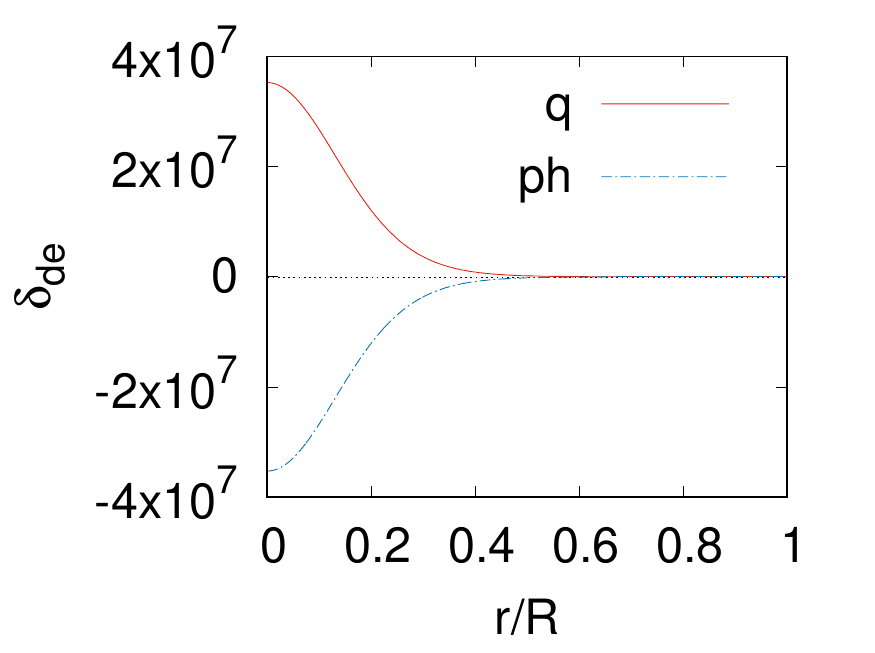}
\centering{c)}
\end{minipage}
\caption{The relative deviation $\delta_{de} (r) = (\rho_{de}(r)-\rho_{\infty})/\rho_{\infty}$ as a function of radial coordinate $r$ inside a WD with radius $R$  for fixed value $x_0=1$ and different values of $c_s$: a) $c_s^2=10^{-3}$; b) $c_s^2=10^{-4}$; c) $c_s^2=10^{-5}$. Solid lines correspond to quintessential dark energy with $\omega_{\infty}=-0.8$, dash-dotted -- phantom dark energy with $\omega_{\infty}=-1.2$.}
\label{fig2}
\end{figure}

Fig.~\ref{fig2} illustrates the calculated relative deviation of density of dark energy from background for different values of effective speed of sound $c_s$ and fixed central density of the matter (or relativistic parameter $x_0$).
It can be seen that with decreasing value of $c_s$ the amount of dark energy inside WD increases for quintessential dark energy (decreases for phantom one) by orders and becomes concentrated towards stellar centre.

\begin{figure}
\begin{minipage}{.32\textwidth}
\includegraphics[width=1\textwidth]{rho_de_r_cs2_1e-4.pdf}
\centering{a)}
\end{minipage}
\begin{minipage}{.32\textwidth}
\includegraphics[width=1\textwidth]{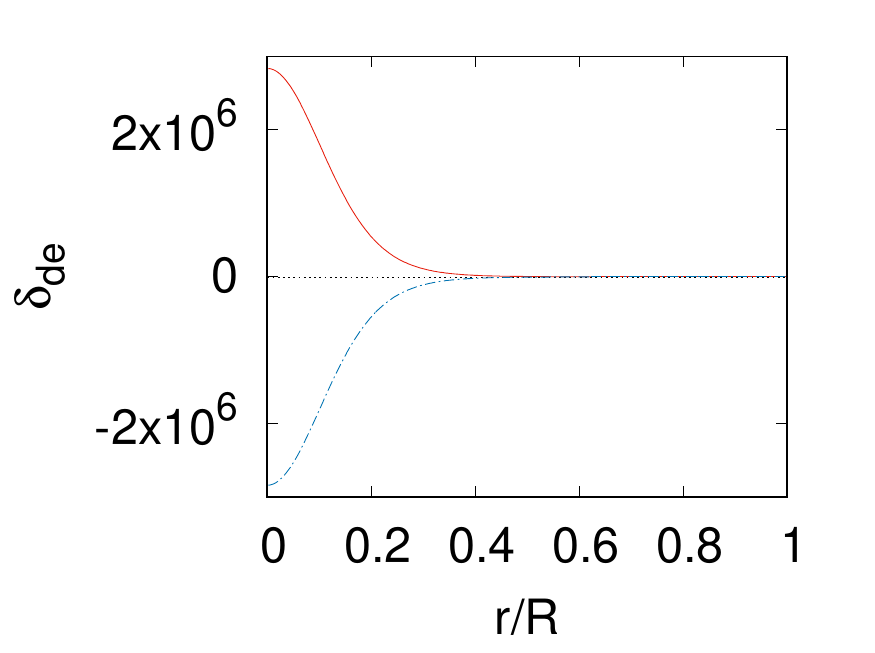}
\centering{b)}
\end{minipage}
\begin{minipage}{.32\textwidth}
\includegraphics[width=1\textwidth]{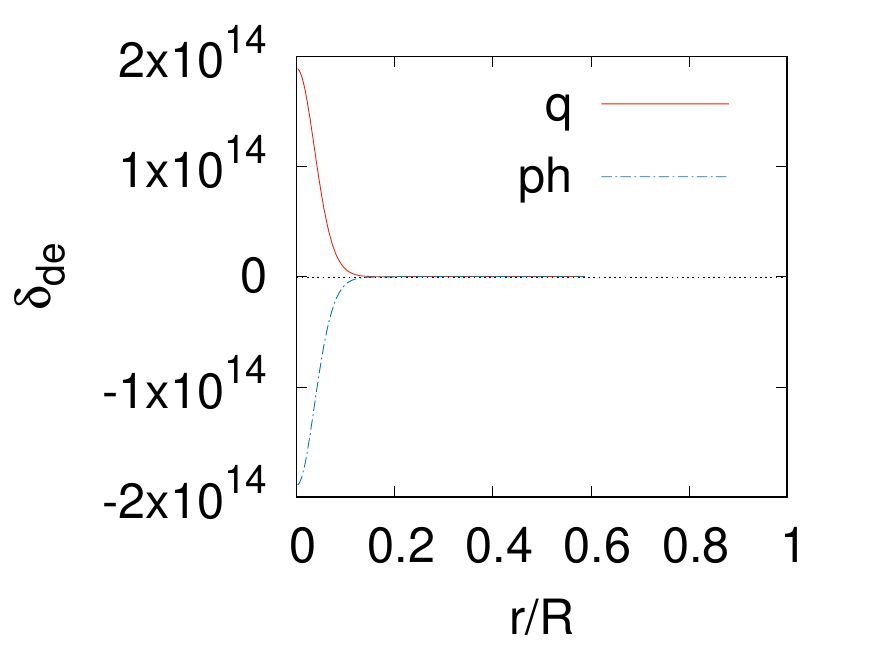}
\centering{c)}
\end{minipage}
\caption{The same as in Fig.~\ref{fig2} but for fixed value $c_s^2=10^{-4}$ and different values of $x_0$: a) $x_0=1$; b) $x_0=5$; c) $x_0=10$.}
\label{fig3}
\end{figure}
Similar behaviour of dark energy one can see when we fixed value of $c_s$ but varied central density (or $x_0$) of a star (see Fig.~\ref{fig3}). Relative change of $\rho_{de}$ is very sensitive to relativistic parameter in stellar centre $x_0$ and is even more abrupt with growing central density of the stellar matter.

\begin{figure}
\centering\includegraphics[width=.8\textwidth]{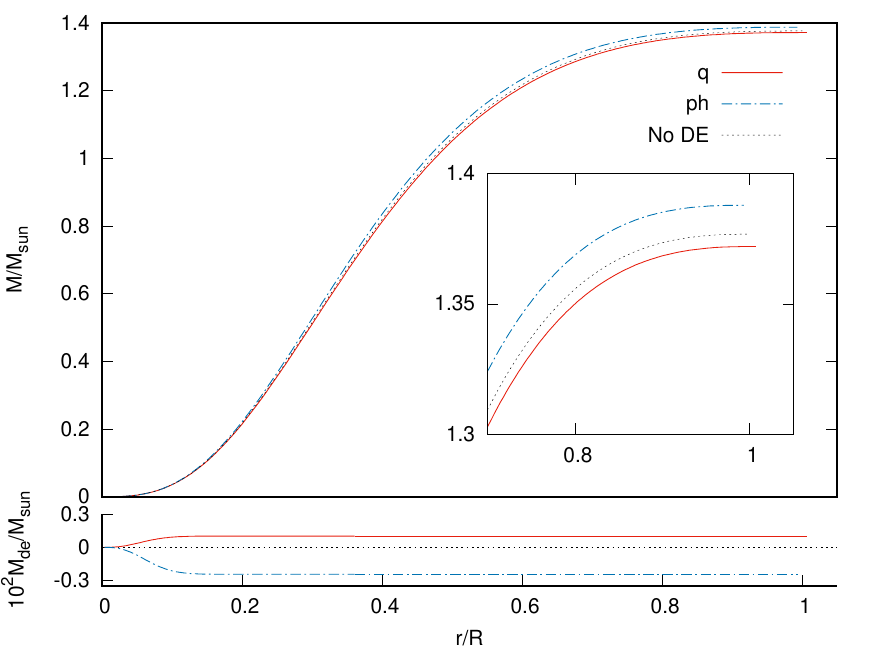}
\caption{Radial dependence of white dwarf mass (top) and mass of dark energy inside a star (bottom). Also, the zoomed-in part of the dependence near stellar surface is shown on the top panel. Solid lines correspond to quintessential dark energy with $\omega_{\infty}=-0.8$, dash-dotted -- phantom one with $\omega_{\infty}=-1.2$. Short-dashed line corresponds to the result of Chandrasekhar model without taking into account dark energy. Here we assumed $x_0=10.2$, $c_s^2=4\cdot 10^{-5}$.}
\label{fig4}
\end{figure}

The radial dependence of WD mass, as well as mass of dark energy inside a star, are shown in figure~\ref{fig4}. It shows that the presence of quintessential dark energy reduces the mass of WD in comparison to result of Chandrasekhar model, whereas phantom dark energy causes its increase. Also, the stellar radius changes pettily, it grows in the case of quintessential dark energy and shrinks in the case of phantom one.

The obtained negative values of density and mass for phantom dark energy needs some comments. The possibility of negative density for this model was already mentioned in \cite{novosyadlyj_2014} (see also \cite{novosyadlyj_2012} for cosmological consequences of phantom models). It was shown in \cite{Carroll2003} and \cite{Cline2004}, that presence of such dark energy can cause UV quantum instability of vacuum through producing a pair of phantom particles and one non-phantom, having energy conserved. The brief discussion of the problem of the existence of physical essence with negative density or mass of particles 
from the general relativity point of view can be found in the recent paper \cite{Socas2019}. Though we consider here classical behaviour of the dark energy and field behind it, this means, that one should be careful when trying to obtain real observable constraints from the solutions for phantom dark energy, remembering there are also quantum limits of such models. In our case, constraints for quintessence and phantom models are almost the same, as divergence occurs almost for the same values of $x_0$ and $c_s^2$.

\section{CONSTRAINT on $c_s$ by WD's MASS-RADIUS RELATION}
\label{sect_4}

\subsection{Influence on mass-radius relation}

\begin{figure*}
\begin{minipage}[t]{0.49\textwidth}
\centering\includegraphics[width=1\textwidth]{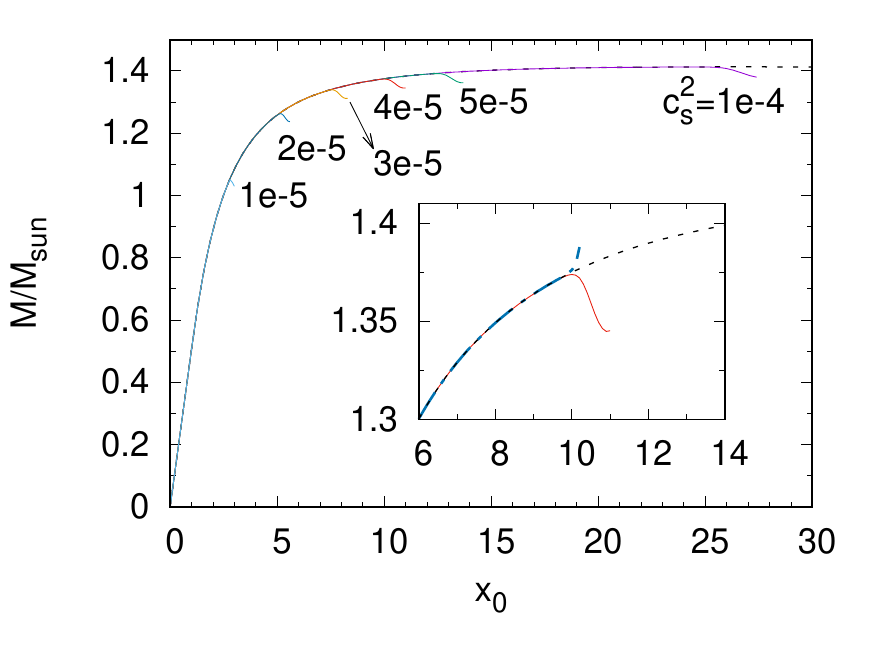}
\end{minipage}
\begin{minipage}[t]{0.49\textwidth}
\centering\includegraphics[width=1\textwidth]{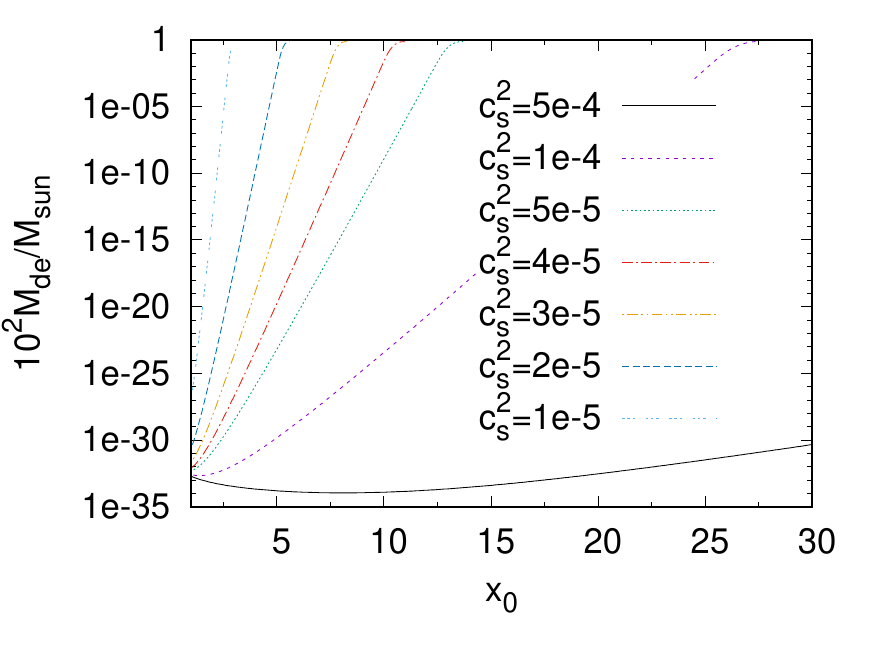}
\end{minipage}
\caption{Left panel: The dependence of WD mass on $x_0$ for a set of values of effective speed of sound of quintessential dark energy $c_s^2$ ($10^{-5}$, $2\cdot 10^{-5}$, $3\cdot 10^{-5}$, $4\cdot 10^{-5}$, $5\cdot 10^{-5}$, $10^{-4}$). Zoomed-in part of the dependence near $x_0=10$ for both types of dark energy with $c_s^2=4\cdot 10^{-5}$ is shown in the central part of the figure. Downward deviation corresponds to quintessential dark energy, upward -- phantom one. Short-dashed line corresponds to the result of Chandrasekhar model. Right panel: The dependence of mass of quintessential type dark energy on parameter $x_0$ for the same set of values for $c_s^2$.}
\label{fig5}
\end{figure*}

We calculated the WD masses and radii for various values of relativistic parameter in stellar centre $x_0$ and effective speed of sound $c_s$. Because we are interested in masses of WDs which can be obtained from observations, our results shown in figure~\ref{fig5} are represented in the form of object's mass as function of $x_0$. 

As we can see from the figure, the dark energy inside WD does not reveal itself unless some critical value of $x_0$ (dependent on $c_s$) is reached. In the vicinity of this value the dark energy accumulated inside a star causes abrupt deviations from the results of Chandrasekhar model (depicted with dotted line). The lower is the value of effective speed of sound, the smaller is critical value of $x_0$ at which the dark energy begins to make a significant contribution into the hydrostatic equilibrium of the WD by lowering its mass in the case of quintessence dark energy and increasing it in the case of phantom one. The reason of such behaviour is following. Higher $x_0$ corresponds to higher density $\rho_m$ of matter in the center of WD and hence, deeper potential well of the system. Deeper potential well causes growth of $\rho_{de}$. In the case of quintessential dark energy, given equation of state of dark energy makes its behaviour close to matter one, which means that for satisfying the hydrostatic equilibrium the lower mass of matter is necessary. Indeed, when $\rho_{de}\gg\bar{\rho}_{de}$ then $w\rightarrow c^2_s$, $p_{de}\rightarrow c^2_s\rho_{de}c^2$ and dark energy contributes into the metric functions, as it follows from equations (\ref{w-rho}) and (\ref{eq3}). This breaks hydrostatic equilibrium and causes gravitational collapse of system. In the case of phantom dark energy $\rho_{de}$ changes its sign and becomes negative. As a consequence, for satisfying the hydrostatic equilibrium the larger mass of matter is necessary. In both cases the process is rapid, as sort of positive feedback loops are created, changing the properties of dark energy inside and in the nearest vicinities of WD.

In the left panel of Fig.~\ref{fig5} the depeneces $M(x_0)$ are presented for WDs in the models without dark energy (short-dashed line) and with quintessence dark energy with different values of $c^2_s$ (main part) and with phantom and quintessence dark energy in the insert which is zoom-in of central part of figure. One can see that deviations for both quintessential and phantom dark energy take place approximately at the same values of $x_0$.

The corresponding masses of dark energy of quintessential type (in log-scale) as a functions of relativistic parameter in stellar centre $x_0$ are shown in the right panel of figure~\ref{fig5}.
The amount of dark energy inside a dwarf steeply increases with $x_0$ and strongly depends on the effective speed of sound $c_s$. Our solutions yield infinite values of densities when $c_s=0$. This is consequence of considered equation of state of the dark energy and, correspondingly, solution (\ref{eq5}), where $c_s$ is in the denominator of exponent. One can conclude, that dark energy with $c_s=0$ is excluded from possible models. 

\subsection{Constraints}

As we saw in Fig.~\ref{fig5}, there are some critical values of $x_0$ depending on $c_s$ at which the dark energy changes the $M-x_0$ relation for WD  noticable. These values are given in Table~\ref{tab1} for both considered types of dark energy.

\begin{table}
\caption{The critical values of relativistic parameter in stellar centre $x_0$ for different values of effective speed of sound for dark energy of both considered types.}
\label{tab1}
\begin{center}
\begin{tabular}{ccc}
\hline\hline
$c_s^2$ & $x_0^q$ & $x_0^{ph}$\\
\hline\\
$10^{-4}$& $27.4$& $25.9$\\
$5\cdot 10^{-5}$& $13.7$& $12.7$\\
$4\cdot 10^{-5}$& $11.0$& $10.2$\\
$3\cdot 10^{-5}$& $8.3$& $7.6$\\
$2\cdot 10^{-5}$& $5.6$& $5.2$\\
$10^{-5}$& $3.0$& $2.8$\\
\hline
\end{tabular}
\end{center}
\end{table}

\begin{figure*}
\begin{minipage}[t]{0.49\textwidth}
\centering\includegraphics[width=1\textwidth]{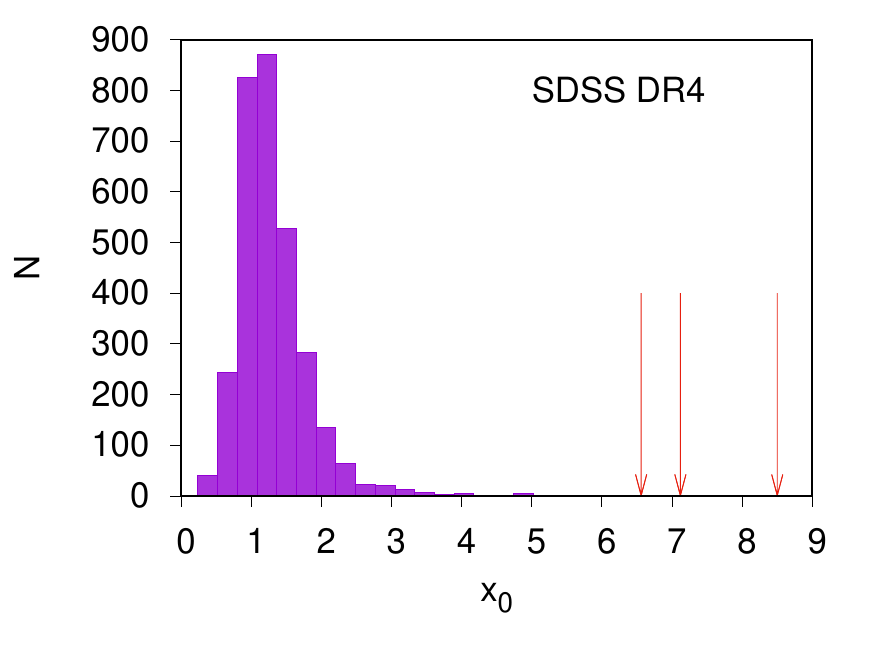}
\end{minipage}
\begin{minipage}[t]{0.49\textwidth}
\centering\includegraphics[width=1\textwidth]{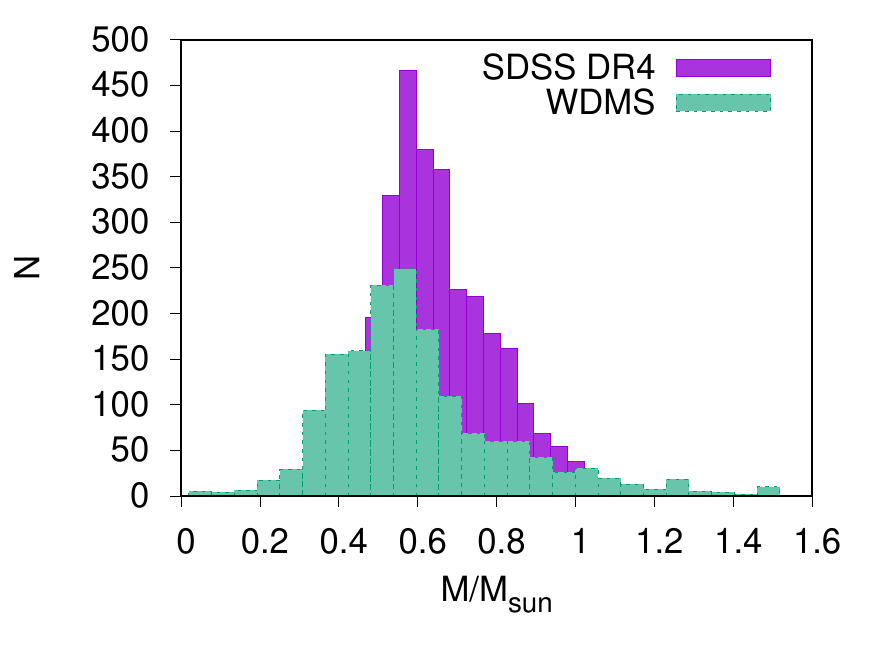}
\end{minipage}
\caption{Left panel: Distribution of white dwarfs of spectral class DA from SDSS DR4 \cite{tremblay_2011} by $x_0$. Right panel: Comparison of distributions by mass of white dwarfs from SDSS DR4 and white dwarfs in binary systems with main sequence stars (WDMS) \cite{rebassa_2010}.}
\label{fig6}
\end{figure*}

In the papers \cite{vavrukh_2011, vavrukh_2012} authors employed the Chandrasekhar model of WDs to solve the inverse problem for large sample ($\sim 3000$) of spectroscopically confirmed WDs of type DA from SDSS DR4 \cite{tremblay_2011}. Using known values of masses and radii there was found that relativistic parameter in stellar centre in the frame of Chandrasekhar model $x_0 \lesssim 2.5$ (left panel of Fig.~\ref{fig6}) for the vast majority of WDs from the sample. However, there exist the outliers for which $x_0$ can be sufficiently high -- up to $x_0^{max}\approx 8.5$ (indicated by arrows in figure). This value can be treated as the maximal value of $x_0$ for real WDs.

Also, WDs in binary systems can be used to estimate the maximal value of $x_0$. As can be seen in the right panel of Fig.~\ref{fig6}, both tails of the distribution by mass (or by $x_0$) are more
populated than for field (single) dwarfs. The reason is a mass transfer between the components of binary systems.
As was mentioned above, it is believed that progenitors of Ia type Super Novae (SN) events are WDs in close binaries with masses near the Chandrasekhar limit. Formally, in the frame of Chandrasekhar model, they occur at $x_0 \rightarrow \infty$.
However, it was shown first in \cite{zeldovich_1971} that mass accretion onto WD
can cause the instability before reaching the Chandrasekhar limit
due to effects of general theory of relativity and/or neutronization.
The critical values of central density in the case of carbon WD were found to be of the same order for both effects ($2.65\cdot 10^{13}$ and $3.90\cdot 10^{13}$~kg/m$^3$, respectively) \cite{shapiro_1983}. The recent values for general theory of relativity effects are very similar (see, for example \cite{mathew_2017}). The corresponding values of relativistic parameter $x_0^{max}$ are $23.8$ and $27.1$, meaning $x_0$ can not exceed it as the SN explosion occures.

Thus, supposing that there exist field WDs with such high masses that correspond up to $x_0^{max}\approx 10$ and/or that Ia type Super Nova events are explosions of WDs in the binaries with relativistic parameter in their centres
$x_0^{max}\approx 25$, we can conclude that the dark energy inside this objects did not reveal itself.
Having limits on $x_0$, one can constrain $c_s^2$. For each value of the former one there is a corresponding value of $c_s^2$ for which the solution for hydrostatical equilibrium equation ceases to exist, i.e. DE with lower values of $c_s^2$ would distroy the WD.
Within this assumption we have found the lower limit for the value of effective speed of sound when the deviation from the Chandrasekhar model becomes non-negligible:
for field (single) WDs $c_s^2\gtrsim 4\cdot 10^{-5}$
and for WDs in binary systems $c_s^2\gtrsim 10^{-4}$. It is interesting to point out, that these constrains are close to the ones, obtained in work \cite{tsizh14}, where we've used current precision of the measuring gravitating mass in the Solar System to constrain the value of the speed of sound of the dark energy.

\section{Conclusions}
\label{sect_5}

In this work we have considered a dark energy with the barotropic equation of state
in static gravitational field of WDs. We have obtained the distribution of dark energy in a WD using Chandrasekhar model and calculated its impact on object's
characteristics in self-consistent way. Investigation of ``mass-radius''
relation for WDs with dark energy inside has shown that deviation of the WD mass from one in the model without dark energy appears to be tiny, unless some critical value of relativistic parameter $x_0$ is reached, though deviation of density of dark energy in the center of star from background dark energy density can be noticeable. Deviation of mass of WD in comparison to Chandrasekhar model is negative for quintessential type of dark energy and positive for phantom one. The critical value of $x_0$ decreases with decreasing value of dark energy effective speed of sound $c_s$.

Using this, we have compared the critical values of relativistic parameter when the concentration of dark energy is too high to maintain the equilibrium of WD with maximum value of $x_0$ obtained in Chandrasekhar model for
observed (single) WDs $x_0^{max}\lesssim 10$ and with value at which another effects affect stellar structure such as neutronization or effects of general theory of relativity for massive WDs in binary systems, $x_0^{max}\lesssim 25$. Supposing that the dark energy has no or has negligible influence on WD structure, which allows WD with such relativistic parameters to exist, we can conclude that minimal value of squared effective speed of sound is $c_s^2\approx 4\cdot 10^{-5}$ in the case of field WDs and $c_s^2\approx 10^{-4}$ for dwarfs near the Chandrasekhar limit in the binary systems.

\section*{Acknowledgements}

This work was supported by the projects of Ministry of Education and Science of Ukraine $\Phi\Phi$-63Hp (No. 0117U007190) and “Formation and characteristics of elements of the structure of the multicomponent Universe, gamma radiation of supernova remnants and observations of variable stars” (No. 0119U001544).

\end{document}